# Nanopore DNA Sequencing Technology: A Sociological Perspective

Suadath V[a*], Muhammad Sajeer P[b*%]

[a] Department of Sociology, Centre for Economic and Social Studies, Hyderabad

[b] Center for Nanoscience and Engineering, Indian Institute of Science, Bangalore

*Both authors contributed equally

%Corresponding authors: muhammads@iisc.ac.in

Abstract

Nanopore sequencing, a next generation sequencing technology, holds the potential to revolutionize multiple facets of life sciences, forensics, and healthcare. While previous research has focused on its technical intricacies and biomedical applications, this paper offers a unique perspective by scrutinizing the societal dimensions (ethical, legal, and social implications) of nanopore sequencing. Employing the lenses of Diffusion and Action Network Theory, we examine the dissemination of nanopore sequencing in society as a potential consumer product, contributing to the field of the sociology of technology. We investigate the possibility of interactions between human and nonhuman actors in developing nanopore technology to analyse how various stakeholders, such as companies, regulators, and researchers, shape the trajectory of the growth of nanopore sequencing. This work offers insights into the social construction of nanopore sequencing, shedding light on the actors, power dynamics, and socio-technical networks that shape its adoption and societal impact. Understanding the sociological dimensions of this transformative technology is vital for responsible development, equitable distribution, and inclusive integration into diverse societal contexts.

1. Introduction

Nanopore-based sequencing technology is rapidly evolving into a pivotal tool with a wide range of applications. These include clinical disease diagnosis, point-of-care testing, monitoring infectious pathogens, and enhancing biology education[1–3]. The fundamental principle of nanopore technology involves monitoring characteristic variations in ionic current as the analyte translocates through a nanoscale pore (1-100 nm in size). A prominent industry player in the field of nanopore sequencing technology is Oxford Nanopore Technologies (ONT), which developed MinION (https://nanoporetech.com/products/sequence/minion), a compact nanopore DNA sequencer with a size comparable to a mobile phone [Figure 1]. ONT commercialized it in 2015, aiming to empower users with the ability to analyze anything, anywhere, and anytime. The technique's affordability and portability have garnered significant attention within the scientific community, facilitating its ongoing transition from a research tool to a consumer-level product. Currently, ONT has developed an even smaller device called SmidgION (https://nanoporetech.com/products/sequence/smidgion) which is designed to be used with a smartphone [Figure 2B]. Remarkably, by introducing a diverse array of nanopore-related products [Figure 2], the company reported a revenue of 133.7 million pounds in 2021 and has been listed on the London Stock Exchange, with an IPO valued at around 5 billion pounds[4].

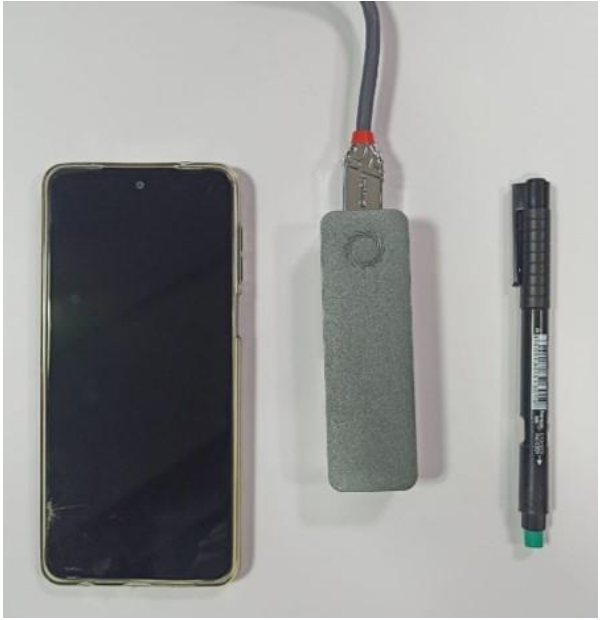

Figure 1: Size comparison of MinION nanopore sequencer with a mobile phone and marker pen.

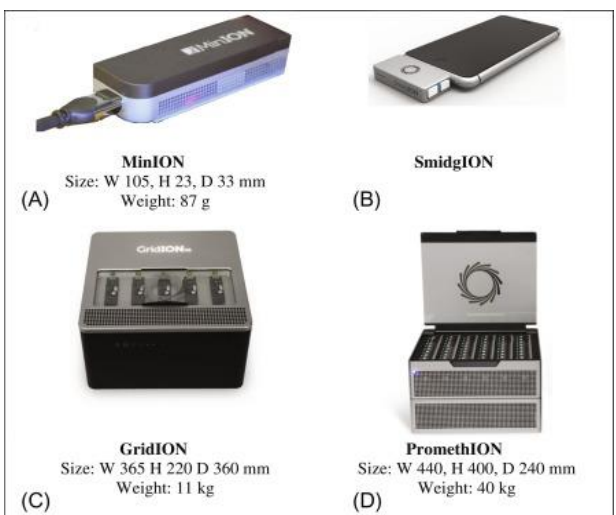

Figure 2: Different sequencing products from Oxford Nanopore Technology. Figure reused with permission from[5]

Nanopore sequencing technology presents a transformative potential across various sectors, from healthcare to agriculture, due to its undeniable technical capabilities. Positive applications have already been demonstrated, ranging from cancer classification to advancements in biology education[6,7]. However, it's crucial to recognize that not all emerging technologies inherently translate into social good. Broader analyses of societal issues regarding emerging technologies are essential to communicate them with the public and various stakeholders simultaneously. Despite numerous articles showcasing the positive technological abilities of nanopore sequencers, there has been a noticeable dearth of exploration into the humanities and social science aspects of nanopore technology. An earlier article by one of the authors provided an introductory exploration into the social science dimension of nanopore sequencing[8].

In this article, we delve into the sociology of nanopore sequencing (ethical, legal, and social implications), aiming to shed light on the social forces likely to influence public acceptance and utilization of this technology. A sociological lens reveals assumptions requiring interrogation about the tech's trajectory and implications. Questions around access, discrimination, values, and regulation deserve deliberation in an inclusive public discourse. By invoking the arguments of diffusion and action network theory, this article will be looking into the possibility of interactions between human and nonhuman actors in the development of this technology to analyze how various stakeholders, such as companies, regulators, and researchers, shape nanopore sequencing. This framework can also help us understand how networks of influence form around technologies and how technologies can have politics embedded in them. Thus, examining the emerging ecosystem around nanopore sequences and their impact on adoption can enable us to interrogate assumptions built into nanopore sequencing and its potential uses.

2. <u>Diffusion theory of innovations and adoption of nanopore sequencing technology</u>

The Diffusion of Innovation theory, initially developed by E.M. Rogers in 1962, offers valuable insights into the potential widespread adoption of nanopore sequencing technology within society. This theory identifies five key characteristics that can accelerate the adoption of an innovation, as given in Table 1.

| Characteristics | Definition | Alignment with MinION from ONT (Y/N) |
|---|---|---|
| Compatibility | The extent to which an innovation aligns with the needs of potential users | Y |
| Trialability | Whether users have the option to test the innovation before committing to its adoption | Y |
| Relative Advantage | How much better is the innovation compared to existing products or methods? | Y |
| Observability | The extent to which the innovation provides visible and compelling results | Y |
| Simplicity/Complexity | The ease or difficulty of using the innovation | Y |

Table 1: Characteristics of diffusion theory of innovations[9] and alignment with nanopore technology

If we evaluate these characteristics in the context of nanopore sequencers, we find that Oxford Nanopore Technologies (ONT) has effectively addressed each one:

- Trialability: ONT introduced a beta program that allowed early users to test the technology, fulfilling the trialability requirement. The portability of devices also helps others to try it easily.
- Compatibility: With features like smaller DNA sample requirements, portability, affordability, and the capacity for long-read sequencing, nanopore sequencers align well with the needs of potential users.
- Relative advantage: Nanopore sequencers, such as the MinION from ONT, offer distinct advantages (similar to the one listed above under compatibility) over other products, positioning them favourably.
- Simplicity/Complexity: Utilising a simple technique based on the coulter counter methodology, nanopore sequencers are user-friendly, even for students.
- Observability: Annual events from ONT, such as London Calling (https://nanoporetech.com/lc24) and recent high-impact publications[10–12] have demonstrated the technology's potential, increasing its observability.

In summary, the MinION and other models of nanopore sequencers successfully align with the characteristics that promote rapid diffusion of innovations, making them well-positioned for broad adoption from an innovation perspective. Nevertheless, it is crucial to examine various societal factors that could influence the widespread adoption of a technology that is capable of significantly updating our understanding of human biology and both genetic and acquired diseases. This demands an analysis from a social science perspective.

### 3. Technological determinism vs Social shaping of technology

Nanopore sequencing offer the potential to make genomic insights more accessible and affordable through decentralized infrastructure, a feature that interests sociologists. However, they emphasize that technology doesn't develop independently of social contexts. Its diffusion is negotiated by relevant groups with various interests, and the current research often overlooks these complexities by focusing narrowly on technical capabilities and individual data privacy. This view aligns with technological determinism, the idea that consumer adoption naturally follows technical advances. But sociologists argue that technologies emerge within complex social systems that mediate their real-world impacts. Public attitudes cannot be assumed, nor can we accept the benefits claimed for such technologies uncritically. Though the regulations offer guidance, they cannot anticipate all the consequences of new technologies.

So, the analysis must move beyond individual ethics to address broader structural issues, such as power dynamics, cultural values, and economic disparities. Scholars like Langdon Winner and Bruno Latour critique the simplistic view of technology. Technological determinism advocates that advancements like Nanopore sequencing will automatically drive societal changes, such as democratizing healthcare or improving research efficiency. However, Winner's theory of technological politics highlights that technologies often reinforce existing power structures, potentially restricting access for marginalized groups despite claims of increased accessibility.

The social shaping of technology theory goes further, stating that technologies are co-constructed by various social actors, including scientists, policymakers, businesses, and the public. Therefore, adoption of nanopore sequencing will depend not only on its technical features but also on who controls its development and use. Its real-world application will be influenced by public discourse, regulatory frameworks, and cultural attitudes. Latour's actor-network theory also reminds us that technologies, like nanopore sequencing, are clearly embedded in networks where human agency, institutional interests, and material devices interact to define their accessibility and usage.

### 4. Access, Inequality, and Discrimination

Nanopore sequencing has the potential to either disrupt or reinforce societal power structures. Despite its decentralised nature, it could still generate disparities in access, favouring the wealthiest individuals, institutions, and nations which could lead to the exclusion of marginalised groups from its benefits, reinforcing existing class-based inequalities. Wealthy institutions and nations may gain substantial advantages from this technology, by leaving those in lower socioeconomic classes without access.

Factors like education, technological awareness, and social mobility could further broaden these disparities, escalating the technological divide.

Nanopore technology can revolutionise healthcare by enabling localised genomic testing, but if only the wealthy can afford it, existing socio-economic inequalities may be exacerbated. By invoking Pierre Bourdieu's theory of capital, wealthier institutions have more economic and cultural capital to acquire and utilise the technology and marginalised groups lack the resources to benefit from it. So, on a global scale, wealthier nations may dominate the access to genomic tools, leading to data monopolies and reinforcing the patterns of neo-colonial exploitation, where developing countries contribute genetic material but do not benefit from the technology.

The major concern on health disparities is in regard to biassed datasets as it could lead to the misinterpretation of minority populations' health needs. Genetic data can also be misused for discrimination, such as linking it to insurance or employment opportunities. This raises the risk of stigmatisation and further marginalisation in the society. Even though the legal protections may help, collective action is essential to prevent unjust treatment based on genetic risks. Nanopore sequencing devices are compact and portable with the potential to be more widely adopted, especially in healthcare. However, the process of adoption may be slow due to costs, technological complexity, and psychological resistance of the public to new methods. Marginalised communities and individuals who have limited technological awareness may be left behind, similar to how some still use outdated technology despite newer advancements or adding to the existing digital divide.

The potential of nanopore sequencing to become a tool of bio power, as conceptualised by Michel Foucault, is highly concerning. Governments or corporations could use the genetic information for surveillance and control. In a panopticon-like society it can further lead to where individuals are categorised and treated based on their genomic data. This could hold serious implications for personal autonomy, freedom, and human rights, particularly for marginalised groups already subject to inordinate state surveillance. In countries with histories of eugenics or caste-based discrimination, the extensive use of genomic technologies with no strong ethical safeguards could lead to new forms of exclusion and oppression. Thus, careful consideration of the ethical and sociological implications of nanopore sequencing is censorious to prevent the misuse and ensure equitable access.

5. <u>Ethical implications and Public engagements</u>

The implementation of nanopore sequencing has significant ethical challenges, particularly around public engagement, informed consent, and data ownership. Thus, we have to stress the need for inclusive dialogue, involving scientists, policymakers, and the public in the decision-making. Advancements of technologies like this should not be introduced with no significant input from those they will affect, ensuring a democratic process that considers ethical concerns and potential risks.

Informed consent must go beyond signing a document; individuals should fully understand how their genetic data will be used and who has access to it. Legal and ethical frameworks often lag behind technological progress, thus clear guidelines are essential for nanopore sequencing, especially as it becomes more popular. Another concern is the structural inequalities. Many low-income countries already lack the infrastructure to integrate such technologies into healthcare systems, exacerbating disparities. So, the possibility of wealthier nations dominating technology, leaving poorer nations dependent even in the future is high. Global health disparities could be exacerbated if diseases more common in affluent regions eclipse those in low income regions.

The potential benefits of nanopore sequencing in healthcare are significant. It could revolutionize diagnostics and reduce dependence on traditional methods. However, this may raise concerns related to financial stability, cultural values, and other societal factors, especially if populations are compelled to adapt.

Governments are increasingly consolidating personal data, raising concerns about integrating genetic information into broader datasets. While there are benefits, such as improved public health, the risks include state-sponsored abuse, such as discrimination or control over reproductive decisions. Without sufficient public engagement and transparency, introducing this technology could provoke protests from vulnerable communities.

Nanopore sequencing also enhances law enforcement capabilities. The technology allows police to analyze DNA samples on-site, reducing the time needed to send samples to forensic labs. However, accuracy in data interpretation is crucial, as mistakes could lead to costly or incorrect conclusions. The Criminal Procedure (Identification) Act, 2022 in India is an example of how this technology is being integrated into policing, but careful regulation is required to avoid misuse.

Consent is vital in all sectors, especially concerning data ownership and privacy. Individuals must understand how their genetic data will be stored, shared, and used for research or clinical purposes. Questions around de-identification, data retention, and secondary analysis must be addressed to ensure transparency and protect privacy.

There are also concerns about the potential misuse of genetic data by employers or insurance companies. Such data could impact employment or insurance claims based on predictive health conditions, infringing on privacy and leading to discrimination. Exploring ancestry through genomic data may also have unintended consequences, such as reinforcing stereotypes or societal stigmas. Introducing nanopore sequencing into societies already dealing with inequality may further complicate existing issues.

Establishing comprehensive regulations and guidelines for data sharing, consent, and collaboration is essential. Institutional review boards can play a critical role in overseeing the ethical use of this technology and ensuring accountability where necessary.

6. <u>Future prospects and recommendations</u>

Nanopore sequencing technology, with its portability, flexibility, and low cost, has great potential for the future of genomics. However, its success and societal acceptance require more than just technological advantages. It is essential to address its ethical implications, ensuring the protection of genetic information and privacy. Understanding the cultural sensitivity of data and developing comprehensive ethical guidelines are key steps toward responsible adoption. Providing training and resources to underrepresented groups can help bridge the technological divide and ensure broader inclusion.

To maintain data control, regulatory agencies must be established to hold researchers accountable, ensure clear communication, and safeguard privacy. Adaptive regulations, such as regulatory sandboxes, should evolve alongside the technology's rapid development. Global partnerships between economically diverse nations can prevent the concentration of technological power. Additionally, it is important to assess the societal impact of sequencing technology and ensure its adoption considers non-medical practices and health customs in different societies.

Governments and institutions must work toward providing access to nanopore technology through funding, partnerships, and skill-enhancing workshops to reduce research disparities across countries. By fostering community engagement, particularly with marginalized groups, specific needs and concerns can be addressed.

While ethical, legal, and social issues surrounding genomics are longstanding, the consumer-level availability of nanopore technology introduces new challenges. Interdisciplinary collaboration between engineering, science, and social sciences can guide the responsible advancement of this technology while upholding democratic and humanitarian values. For nanopore sequencing to realize its healthcare potential, its adoption must prioritize social justice. Training programs, equitable funding, and international collaboration are essential to ensuring the technology benefits all sectors of society while avoiding new forms of exclusion.

Thus, sociologists advocate for a reflexive approach to adopting technology, which emphasises the ongoing reassessment of social impacts and the flexibility to revise policies as new challenges arise. In conclusion, while nanopore sequencing offers exciting advancements in genomics and healthcare, its implementation requires careful consideration of social, ethical, and political factors to ensure equitable distribution of benefits and the responsible management of risks.

6. <u>Acknowledgments</u>

We extend our sincere appreciation to Prof. Manoj Varma (IISc Bangalore), Dr. CH Krishna Rao (Centre for Economic and Social Studies, Hyderabad), Arun Kumar PK, Jaise Johnson, and Anumol Dominic for the invaluable feedback and constructive suggestions provided. Their thoughtful input has been instrumental in improving the quality of our work. Muhammad Sajeer P also acknowledges the PhD fellowship from the Government of India. Muhammad Sajeer P also thanks the members of the nanopore group at IISc Bangalore for being supportive and helpful.

7. <u>Data availability</u>

There was no data related to this manuscript

8. <u>Reference</u>